\documentclass[12pt,preprint]{aastex}

\newcommand{\be}{\begin{equation}}
\newcommand{\ee}{\end{equation}}
\newcommand{\bea}{\begin{eqnarray}}
\newcommand{\eea}{\end{eqnarray}}

\begin{document}

\title{COSMIC RAY SCATTERING BY COMPRESSIBLE MAGNETOHYDRODYNAMIC TURBULENCE}

\author{Huirong Yan and A. Lazarian}

\date{\today{}}

\begin{abstract}
Recent advances in understanding of magnetohydrodynamic (MHD) turbulence
call for substantial revisions in the picture of cosmic ray transport.
In this paper we use recently obtained scaling laws for MHD modes
to calculate the scattering frequency for cosmic rays in the ISM.
We consider gyroresonance with MHD modes (Alfv\'{e}nic, slow and
fast) and transit-time damping (TTD) by fast modes. We provide calculations
of cosmic ray scattering for various phases of interstellar medium
with realistic interstellar turbulence driving that is consistent
with the velocity dispersions observed in diffuse gas. We account
for the turbulence cutoff arising from both collisional and collisionless
damping. We obtain analytical expressions for diffusion coefficients that enter Fokker-Planck equation describing cosmic ray evolution. We calculate the scattering rate and parallel spatial diffusion coefficients of cosmic rays for both Alfv\'{e}nic
and fast modes. We conclude that fast modes
provides the dominant contribution to cosmic ray scattering for the
typical interstellar conditions in spite of the fact that fast modes
are subjected to damping. We show that the efficiency of the scattering
depends on the plasma $\beta$ since it determines the damping of
the fast modes. We also show that the streaming instability is modified in the presence of turbulence. 
\end{abstract}


\section{Introduction}

Most astrophysical systems, e.g. accretion disks, stellar winds, the
interstellar medium (ISM) and intercluster medium are turbulent with
an embedded magnetic field that influences almost all of their properties.
High conductivity of the astrophysical fluids makes the magnetic fields
{}``frozen in'', and influence fluid motions. The coupled motion
of magnetic field and conducting fluid holds the key to many astrophysical
processes. 

The propagation of cosmic rays (CRs) is affected by their interaction
with magnetic field. This field is turbulent and therefore, the resonant
interaction of cosmic rays with MHD turbulence has been discussed
by many authors as the principal mechanism to scatter and isotropize
cosmic rays (Schlickeiser 2002). Although cosmic ray diffusion can
happen while cosmic rays follow wandering magnetic fields (Jokipii
1966), the acceleration of cosmic rays requires efficient scattering.
For instance, scattering of cosmic rays back into the shock is a
vital component of the first order Fermi acceleration (see Longair
1997).

While most investigations are restricted to Alfv\'{e}n modes propagating
along an external magnetic field (the so-called slab model of Alfv\'{e}nic
turbulence), obliquely propagating MHD modes have been included in
Fisk et al. (1974) and later studies (Bieber et al. 1988, Pryadko
\& Petrosian 2000). The problem, however, is that the Alfv\'{e}nic
turbulence considered in their studies is isotropic turbulence, and
this contradicts to the modern understanding of MHD turbulence (Goldreich
\& Shridhar 1995, see Cho, Lazarian \& Yan 2002 for a review and references
therein).

A recent study (Lerche \& Schlickeiser 2001) found a strong dependence
of scattering on turbulence anisotropy. Therefore the calculations
of CR scattering must be done using a realistic MHD turbulence model.
An important attempt in this direction was carried out in Chandran
(2000). However, only incompressible motions were considered. On the contrary, ISM
is highly compressible. Compressible MHD turbulence has been studied
recently (see review by Cho \& Lazarian 2003a and references therein).
Schlickeiser \& Miller (1998) addressed the scattering by fast modes.
But they did not consider the damping, which is essential for fast
modes. In this paper we discuss in detail the various damping processes
which can affect the fast modes. To characterize the turbulence we
use the statistics of Alfv\'{e}nic modes obtained in Cho, Lazarian
\& Vishniac (2002, henceforth CLV02) and compressible modes obtained
in Cho \& Lazarian (2002, henceforth CL02, 2003b,c).

Yan \& Lazarian (2002) used recent advances in understanding of MHD
turbulence (CL02) to describe cosmic ray propagation in the galactic
halo. In this paper we undertake a comprehensive study of cosmic ray scattering rates. In $\S$2,
we describe the statistics of Alfv\'{e}nic and compressible turbulence
in various conditions, including both high $\beta$ and low $\beta$
cases. In $\S$3, we describe the resonant interactions between the
MHD modes and CRs. The scattering
by Alfv\'{e}n modes and fast modes is presented in $\S$4. We
apply our results to different phases of ISM, including Galactic halo,
hot ionized medium (HIM), warm ionized medium (WIM) and also partially
ionized medium. In $\S$5, we describe the streaming instability in
the presence of background turbulence. Discussion of our results is provided in $\S$6 while the summary is given in $\S$7.

\section{MHD cascade and its damping}

\subsection{MHD turbulence cascade}

A substantial shortcoming of earlier studies was that the crucial element for cosmic ray scattering, namely, MHD turbulence model was taken somewhat ad hoc. The goal of the current paper is to use the recent advances in quantitative description of MHD turbulence to quantify scattering of cosmic rays.

MHD perturbations can be decomposed into Alfv\'{e}nic, slow and fast
waves with well-known dispersion relations (see Alfv\'{e}n \& F\"{a}lthmmar 1963). Alfv\'{e}nic turbulence
is considered by many authors as the default model of interstellar
magnetic turbulence. This is partially motivated by the fact that
unlike compressible modes, the Alfv\'{e}n ones are essentially free
of damping in fully ionized medium (see Ginzburg 1961, Kulsrud \&
Pearce 1969).  Important questions arise. Can the MHD perturbations that characterize turbulence be separated into distinct modes? Can the linear modes be used for this purpose? The separation into Alfv\'en and pseudo-Alfv\'en modes is the cornerstone of the Goldreich-Sridhar (1995, henceforth GS95) model of turbulence. This model and the legitimacy of the separation were tested successfully with numerical simulations (Cho \& Vishniac 2000; Maron \& Goldreich 2001, CLV02). Separation of MHD perturbations in compressible media into fast, slow and Alfv\'en modes is discussed in GS95, Lithwick \& Goldreich 2001, CL02). The actual decomposition of MHD turbulence into Alfv\'en, slow and fast modes was performed in CL02, Cho \& Lazarian
(2003, henceforth CL03), who also quantified the intensity of the interaction between different modes (see below).

Turbulence of Alfv\'en modes is the default model of MHD turbulence for many resources. Other models are considered less important due to damping. However, within an earlier research it was frequently forgotten that, unlike hydrodynamic
turbulence, Alfv\'{e}nic turbulence is anisotropic, with eddies elongated
along the magnetic field (see Higdon 1984, Shebalin et al 1983). This
happens because it is easier to mix the magnetic field lines perpendicular
to the direction of the magnetic field rather than to bend them. The
GS95 model describes \textit{incompressible} Alfv\'{e}nic turbulence,
which formally means that plasma $\beta=P_{gas}/P_{mag}$, the ratio
of gas pressure to magnetic pressure is infinitely large. The corresponding
scaling can be easily obtained. For instance, calculations in CLV02
prove that motions perpendicular to magnetic field lines are essentially
hydrodynamic. As the result, energy transfer rate due to those motions
is constant $\dot{E_{k}}\sim v_{k}^{2}/\tau_{k}$, where $\tau_{k}$
is the energy eddy turnover time $\sim(v_{k}k_{\perp})^{-1}$, where
$k_{\perp}$ is the perpendicular component of the wave vector $\mathbf{k}$.
The mixing motions couple to the wave-like motions parallel to magnetic
field giving a critical balance condition, i.e., $k_{\bot}v_{k}\sim k_{\parallel}V_{A}$,
where $k_{\parallel}$ is the parallel component of the wave vector
$\mathbf{k}$, $V_{A}$ is the Alfv\'en speed%
\footnote{note that the linear dispersion relation is used for Alfv\'en modes.%
}. From these arguments,
the scale dependent anisotropy $k_{\parallel}\propto k_{\perp}^{2/3}$and
a Kolmogorov-like spectrum for the perpendicular motions $v_{k}\propto k^{-1/3}$
can be obtained (see Lazarian \& Vishniac 1999).

It was conjectured in Lithwick \& Goldreich (2001) that GS95 scaling
should be approximately true for Alfv\'{e}n and slow modes in moderately
compressible plasma. For magnetically dominated, the so-called low
$\beta$ plasma, CL02 showed that the coupling of Alfv\'{e}nic and
compressible modes is weak and that the Alfv\'{e}nic and slow modes
follow the GS95 spectrum. This is consistent with the analysis of
HI velocity statistics (Lazarian \& Pogosyan 2000, Stanimirovic \&
Lazarian 2001) as well as with the electron density statistics (see
Armstrong, Rickett \& Spangler 1995). Calculations in  CL03 demonstrated that fast modes are marginally
affected by Alfv\'{e}n modes and follow acoustic cascade in both
high and low $\beta$ medium. In what follows, we consider both Alfv\'{e}n
modes and compressible modes and use the description of those modes
obtained in CL02, CL03 to study CR scattering by MHD turbulence.

The distribution of energy between compressible and incompressible modes depends, in general, on the way turbulence is driven. CL02 and CL03 studied generation of compressible perturbations using random incompressible driving. As the result we obtained an expression that relates the energy in fast $\sim\delta V_f^2$ and Alfv\'en $\sim\delta V_A^2$ modes, 

\be
(\delta V_f/\delta V_A)^2\sim \delta V_A\times V_A/(V_A^2+C_S^2),
\ee
where $C_S$ is the sound speed. This relation testifies that at large scales incompressible driving can transfer an appreciable part of energy into fast modes. However, at smaller scales the drain of energy from Alfv\'en to fast modes is marginal. Therefore the cascades evolve without much of cross talk. Naturally a more systematic study of different types of driving is required. In the absence of this, in what follows we assume that equal amounts of energy are transfered into fast and Alfv\'en modes when driving is at large scales.

At small scales turbulence spectrum is truncated by damping. Various
processes can damp the MHD motions (see Appendix A for details).
In partially ionized plasma, the ion-neutral collisions are the dominant
damping process. In fully ionized plasma, there are basically two
kinds of damping: collisional or collisionless damping. Their relative
importance depends on the mean free path 
in the medium (Braginskii 1965),

\be
l_{mfp}=v_{th}\tau=6\times10^{11}(T/8000K)^{2}/n.\label{meanfp}
\ee
If the wavelength is larger than
the mean free path, viscous damping dominates. If, on the other hand,
the wavelength is smaller than mean free path, then the plasma is
in the collisionless regime and collisionless damping is dominant.

To obtain the truncation scale, the damping time $\Gamma_{d}^{-1}$
should be compared to the cascading time $\tau_{k}$. As we mentioned
earlier, the Alfv\'{e}nic turbulence cascades over one eddy turn
over time $(k_{\perp}v_{k})^{-1}\sim(k_{\parallel}V_{A})^{-1}$. The
cascade of fast modes takes a bit longer: 

\be
\tau_{k}=\omega/k^{2}v_{k}^{2}=(k/L)^{-1/2}\times V_{ph}/V^{2},\label{fdecay}
\ee
where $V$ is the turbulence velocity at the injection scale, $V_{ph}$ is is the phase speed of fast modes and equal to Alfv\'en and sound velocity for high and
low $\beta$ plasma, respectively (CL02). If the damping is faster than
the cascade, the turbulence is truncated. Otherwise, for the sake
of simplicity, we ignore the damping and assume that the turbulence
cascade is unaffected. As the transfer of energy between
Alfv\'en, slow and fast modes of MHD turbulence is suppressed, we consider different components of MHD cascade independently.

We get the cutoff scale $k_{c}$ by equating the damping rate and
cascading rate $\tau_{k}\Gamma_{d}\simeq1$. Then we check whether
it is self-consistent by comparing the $k_{c}$ with the relevant
scales, e.g., injection scale, mean free path and the proton gyro-scale.

Damping is, in general, anisotropic, i.e., the linear damping (see Appendix~A) depends on the
angle between the wavevector ${\bf k}$ and local direction of
magnetic field ${\bf B}$. In the typical ISM
this happens at the sufficiently small
scales where the direction of ${\bf B}$ is well defined (see discussion in the review by Cho, Lazarian \& Vishniac 2003).  

Note that unless randomization of wave vectors is comparable to the cascading rate the damping scale gets angle-dependent. Consider fast
modes, that will be shown to be 
the most important for cosmic ray scattering. Their non-linear 
cascading can be characterized by interacting 
wavevectors ${\bf k}$ that are nearly
collinear (see review by Cho, Lazarian \& Vishniac 2003). The
possible transversal deviation $\delta k$ can be estimated 
from the uncertainty condition $\delta \omega t_{cas}\sim 1$,
where $\delta \omega\sim V_{ph} \delta k (\delta k/k)$. As $t_{cas}\sim 1/k^4 v_k\sim k^{-1/2}$ (see
CL02) the $\delta k$ scales as $k^{3/4}$. Therefore
the deviation in ${\bf k}$ direction in the course of interactions 
decreases with the increase of the wavenumber as $\delta k/k \sim k^{-1/4}$,
and the number of interactions required to randomize ${\bf k}$ increases
as $k^{1/2}$ with the increase of the wavenumber. This means that the
randomization of ${\bf k}$ gets marginal for large $k$. In the presence
of anisotropic damping this results in anisotropic distribution of
fast mode energy at small $k$.

With this input at hand, it is possible to determine the turbulence
damping scales for a given medium. For modes in fully ionized medium
one should compare the wavelength and mean free path before determining
which damping we should apply. However, we are dealing with the turbulence
cascade. Therefore, the situation is more complicated: If the mean
free path is larger than the turbulence injection scale, we can simply
apply the collisionless damping to the whole inertial range of turbulence.
In general one should compare the viscous cut off and the mean free
path. If the cutoff scale is larger than the mean free path, it shows
that the turbulence is indeed cut off by the viscous damping. Otherwise,
we neglect the viscous damping and just apply the collisionless damping
to the turbulence below the mean free path%
\footnote{We shall show that for some angles between ${\bf B}$ and ${\bf k}$ the damping may result in turbulence transferring to collisionless regime only over a limit range of angles.%
}. By comparing different
damping, we find the dominant damping processes for the idealized
ISM phases (see table1).
\begin{table*}
\begin{tabular}{|c|c|c|c|c|c|c|}
\hline 
ISM&
 galactic halo&
 HIM&
 WIM&
 WNM&
 CNM&
 MC\tabularnewline
\hline
T(K)&
 $2\times 10^6$&
 $1\times10^{6}$&
 8000&
 6000&
 100&
 15\tabularnewline
\hline
n(cm$^{-3}$)&
 $10^{-3}$&
 $4\times10^{-3}$&
 0.1&
 0.4&
 30&
 200\tabularnewline
\hline
$l_{mfp}$(cm)&
$4\times 10^{19}$&
$2\times10^{18}$&
$6\times10^{12}$&
$8\times10^{11}$&
$3\times10^{6}$&
$10^{4}$\tabularnewline
\hline
L(pc)&
 100&
 100&
 50&
 50&
 50&
 50\tabularnewline
\hline
B($\mu$G)&
5&
2&
5&
5&
5&
15\tabularnewline
\hline
damping&
 collisionless&
 collisionless&
 collisional&
 neutral-ion&
 neutral-ion&
 neutral-ion\tabularnewline
\end{tabular}

\caption{The parameters of idealized ISM phases and relevant damping. The
dominant damping mechanism for turbulence is given in the last line. HIM=hot ionized medium, CNM=cold neutral medium, WNM=warm neutral
medium, WIM=warm ionized medium, DC=dark cloud.}
\end{table*}

\subsection{Statistics of fluctuations}

Within random-phase approximation, the correlation tensor in Fourier
space is (see Schlickeiser \& Achatz 1993)

\begin{eqnarray}
<B_{i}(\mathbf{k})B_{j}^{*}(\mathbf{k'})>/B_{0}^{2}=\delta(\mathbf{k}-\mathbf{k'})M_{ij}(\mathbf{k}),\nonumber \\
<v_{i}(\mathbf{k})B_{j}^{*}(\mathbf{k'})>/V_{A}B_{0}=\delta(\mathbf{k}-\mathbf{k'})C_{ij}(\mathbf{k}),\nonumber \\
<v_{i}(\mathbf{k})v_{j}^{*}(\mathbf{k'})>/V_{A}^{2}=\delta(\mathbf{k}-\mathbf{k'})K_{ij}(\mathbf{k}),\end{eqnarray}
 where $B_{\alpha,\beta}$, $v_{\alpha,\beta}$ are respectively the
magnetic and velocity perturbation associated with the turbulence.
For the balanced cascade we consider, i.e., equal intensity of forward
and backward modes, $C_{ij}(\mathbf{k})=0$.

The isotropic tensor usually used in the literature is \begin{equation}
M_{ij}(\mathbf{k})=C_{0}\{\delta_{ij}-k_{i}k_{j}/k^{2}\} k^{-11/3},\label{isotropic}\end{equation}
 The normalization constant $C_{0}$ here can be obtained if the energy
input at the scale $L$ is defined. Assuming equipartition $u_{k}=\int dk^{3}\sum_{i=1}^{3}M_{ii}B_{0}^{2}/8\pi\sim B_{0}^{2}/8\pi$,
we get $C_{0}=L^{-2/3}/12\pi$. The normalization for the following
tensors are obtained in the same way.

The analytical fit to the anisotropic tensor for Alfv\'{e}n modes,
obtained in CLV02 is,

\begin{equation}
\left[\begin{array}{c}
M_{ij}({\mathbf{k}})\\
K_{ij}({\mathbf{k}})\end{array}\right]=\frac{L^{-1/3}}{6\pi}I_{ij}k_{\perp}^{-10/3}\exp(-L^{1/3}|k_{\parallel}|/k_{\perp}^{2/3}),\label{anisotropic}\end{equation}
 where $I_{ij}=\{\delta_{ij}-k_{i}k_{j}/k^{2}\}$ is a 2D tensor in
$x-y$ plane which is perpendicular to the magnetic field, $L$ is
the injection scale, $V$ is the velocity at the injection scale.
Slow modes are passive and similar to Alfv\'{e}n modes. 

According to CL02, fast modes are isotropic and have one dimensional
energy spectrum $E(k)\propto k^{-3/2}$. In low $\beta$ medium, the
corresponding correlation is (YL02)\begin{equation}
\left[\begin{array}{c}
M_{ij}({\mathbf{k}})\\
K_{ij}({\mathbf{k}})\end{array}\right]={\frac{L^{-1/2}}{8\pi}}H_{ij}k^{-7/2}\left[\begin{array}{c}
\cos^{2}\theta\\
1\end{array}\right],\label{lbtensor}\end{equation}
 where $\theta$ is the angle between $\mathbf{k}$ and $\mathbf{B}$,
$H_{ij}=k_{i}k_{j}/k_{\perp}^{2}$ is also a 2D tensor in $x-y$ plane.
The factor $\cos^{2}\theta$ represents the projection as magnetic
perturbation is perpendicular to $\mathbf{k}$. This tensor is different
from that in Schlickeiser \& Miller (1998)%
\footnote{Here we give only the x,y component for the perturbation, the solenoidal condition will be satisfied if the z component is added.}. For isotropic turbulence,
the tensor of the form $\propto E_{k}(\delta_{ij}-k_{i}k_{j}/k^{2})$
was obtained to satisfy the condition $\mathbf{k}\cdot\delta\mathbf{B}=0$
(see Schlickeiser 2002). Nevertheless, the fact that $\delta\mathbf{B}$
in fast modes is in the $\mathbf{k}$-$\mathbf{B}$ plane place another
constraint on the tensor so that the term $\delta_{ij}$ doesn't exist.

In high $\beta$ medium, fast modes in this regime are essentially
sound modes compressing magnetic field (GS95,
Lithwick \& Goldreich 2001, CL03). The compression of magnetic field
depends on plasma $\beta$. The corresponding x-y components of the
tensors are \begin{equation}
\left[\begin{array}{c}
M_{ij}({\mathbf{k}})\\
K_{ij}({\mathbf{k}})\end{array}\right]={\frac{L^{-1/2}}{2\pi}}\sin^{2}\theta H_{ij}k^{-7/2}\left[\begin{array}{c}
\cos^{2}\theta/\beta^{2}\\
1/\beta\end{array}\right].\label{hbtensor}\end{equation}
The velocity perturbation in high
$\beta$ medium is radial, i.e., along $\mathbf{k}$, thus we have
the factor $\sin^{2}\theta$ and also $C_{s}^{2}/V_{A}^{2}=2/\beta$
from the magnetic frozen condition $\omega\delta\mathbf{B}\sim\mathbf{k}\times(\mathbf{v}_{k}\times\mathbf{B})$.
In high $\beta$ medium, the energy of fast modes are reduced by $V_{A}^{2}/C_{S}^{2}$,
this gives another $2/\beta$. We use these statistics to calculate
cosmic ray scattering arising from MHD turbulence.

\section{Interactions between turbulence and particles}

Particles get into resonance with MHD perturbations if the resonant
condition is fulfilled, namely, $\omega=k_{\parallel}v\mu+n\Omega$ ($n=\pm1,2...$),
where $\omega$ is the wave frequency, $\Omega=\Omega_{0}/\gamma$
is the gyrofrequency of relativistic particle, $\mu=\cos\xi$,
where $\xi$ is the pitch angle of particles. 

Basically there are
two main types of resonant interactions: gyroresonance acceleration
and transit acceleration (henceforth TTD). The latter requires longitudinal
motions and it only operates with compressible modes. Gyroresonance
occurs when the Doppler shifted frequency of the wave in the particle's
guiding center rest frame $\omega_{gc}=\omega-k_{\parallel}v\mu$
is a multiple of the particle gyrofrequency, and the rotating direction
of wave electric vector is the same with the direction for Larmor
gyration of particle. For high energy particles, the resonance happens
for both positive and negative $n$. TTD happens due to the resonant
interaction with parallel magnetic mirror force $-(mv_{\perp}^{2}/2B)\nabla_{\parallel}\mathbf{B}$.
For small amplitude modes, particles should be in phase with the wave
so as to have a secular interaction with wave (Schlickeiser \& Miller
1998). This gives the Cherenkov resonant condition $\omega-k_{\parallel}v_{\parallel}=0$.

We employ quasi-linear theory (QLT) to obtain our estimates. QLT has
been proved to be a useful tool in spite of its intrinsic limitations
(Schlickeiser \& Miller 1998, Miller 1997). For moderate energy cosmic
rays, the corresponding resonant scales are much smaller than the
injection scale. Therefore the fluctuation on the resonant scale $\delta B\ll B_{0}$
even if they are comparable at the injection scale. QLT disregards
diffusion of cosmic rays that follow wandering magnetic field lines
(Jokipii 1966) and this diffusion should be accounted separately. If mean magnetic field is larger than the fluctuations at the injection scale, we may say that the QLT treatment we employ defines parallel diffusion.
Obtained by applying the QLT to the collisionless Boltzmann-Vlasov
equation, the Fokker-Planck equation is generally used to describe
the evolvement of the gyrophase-averaged distribution function $f$,

\[
\frac{\partial f}{\partial t}=\frac{\partial}{\partial\mu}\left(D_{\mu\mu}\frac{\partial f}{\partial\mu}+D_{\mu p}\frac{\partial f}{\partial p}\right)+\frac{1}{p^{2}}\frac{\partial}{\partial p}\left[p^{2}\left(D_{\mu p}\frac{\partial f}{\partial\mu}+D_{pp}\frac{\partial f}{\partial p}\right)\right],\]
 where $p$ is the particle momentum. The Fokker-Planck coefficients
$D_{\mu\mu},D_{\mu p},D_{pp}$ are the fundamental physical parameter
for measuring the stochastic interactions, which are determined by
the electromagnetic fluctuations (Schlickeiser 1993):

Adopting the approach in (Schlickeiser 1993) and taking into account only the dominant interaction at $n=\pm1$, we can get the Fokker-Planck
coefficients (see
Appendix B),

\begin{eqnarray}
\left(\begin{array}{c}
D_{\mu\mu}\\
D_{pp}\end{array}\right)  =  {\frac{\pi\Omega^{2}(1-\mu^{2})}{2}}\int_{k_{min}}^{k_{max}}dk^3\delta(k_{\parallel}v_{\parallel}-\omega+\pm \Omega)\nonumber\\
\left(\begin{array}{c}
\left(1+\frac{\mu V_{ph}}{v\zeta}\right)^{2}\\
m^{2}V_{A}^{2}\end{array}\right)\left\{ (J_{2}^{2}({\frac{k_{\perp}v_{\perp}}{\Omega}})+J_{0}^{2}({\frac{k_{\perp}v_{\perp}}{\Omega}}))\right.\nonumber\\
\left[\begin{array}{c}
M_{{\mathcal{RR}}}({\mathbf{k}})+M_{{\mathcal{LL}}}({\mathbf{k}})\\
K_{{\mathcal{RR}}}({\mathbf{k}})+K_{{\mathcal{LL}}}({\mathbf{k}})\end{array}\right]-2J_{2}({\frac{k_{\perp}v_{\perp}}{\Omega}})J_{0}({\frac{k_{\perp}v_{\perp}}{\Omega}})\nonumber\\
\left.\left[e^{i2\phi}\left[\begin{array}{c}
M_{{\mathcal{RL}}}({\mathbf{k}})\\
K_{{\mathcal{RL}}}({\mathbf{k}})\end{array}\right]+e^{-i2\phi}\left[\begin{array}{c}
M_{{\mathcal{LR}}}({\mathbf{k}})\\
K_{{\mathcal{LR}}}({\mathbf{k}})\end{array}\right]\right]\right\} ,\label{gyro}
\end{eqnarray}
where $\zeta=1$ for $k_{min}=L^{-1}$, $k_{max}=\Omega_{0}/v_{th}$
corresponds to the dissipation scale, $m=\gamma m_{H}$ is the relativistic
mass of the proton, $v_{\perp}$ is the particle's velocity component
perpendicular to $\mathbf{B}_{0}$, $\phi=\arctan(k_{y}/k_{x}),$
${\mathcal{L}},{\mathcal{R}}=(x\pm iy)/\sqrt{2}$ represent left and
right hand polarization.

The delta function $\delta(k_{\parallel}v_{\parallel}-\omega+n\Omega)$
approximation to real interaction is true when magnetic perturbations
can be considered static%
\footnote{Cosmic rays have such high velocities that the slow variation of the
magnetic field with time can be neglected. %
} (Schlickeiser 1993). For cosmic rays, $k_{\parallel}v_{\parallel}\gg\omega$
so that the resonant condition is just $k_{\parallel}v\mu+n\Omega=0$.
From this resonance condition, we know that the most important interaction
occurs at $k_{\parallel}=k_{\parallel,res}=\Omega/v_{\parallel}$.
This is generally true except for small $\mu$ (or scattering near
$90^{o}$). In this case, these particles move at a speed $v\mu$ close to Alfv\'{e}n
speed along the magnetic field so that we have to take into account
the dynamics of the turbulence (Schlickeiser 1993, Bieber, Matthaeus
\& Smith 1994, Yan \& Lazarian 2003).

\section{Scattering of cosmic rays}

\subsection{Scattering by Alfv\'{e}nic turbulence}

As we discussed in $\S$2, Alfv\'{e}n modes are anisotropic, eddies
are elongated along the magnetic field, i.e., $k_{\perp}>k_{\parallel}$.
The scattering of CRs by Alfv\'{e}n modes is suppressed first because
most turbulent energy goes to $k_{\perp}$ due to the anisotropy of
the Alfv\'{e}nic turbulence so that there is much less energy left
in the resonance point $k_{\parallel,res}=\Omega/v_{\parallel}\sim r_{L}^{-1}$.
Furthermore, $k_{\perp}\gg k_{\parallel}$ means $k_{\perp}\gg r_{L}^{-1}$
so that cosmic ray particles have to be interacting with lots of eddies
in one gyro period. This random walk substantially decreases the scattering
efficiency. Mathematically, this effect is embodied by the Bessel
functions in the Fokker-Plank coefficients. We know that the asymptotic
expression of Bessel function $J_{n}(x)\sim1/\sqrt{x}$. So the suppression
will be of the order $(\Omega/k_{\perp}v_{\perp})^{\frac{1}{2}}=(k_{\perp}r_{L})^{\frac{1}{2}}$.
Noticing that $k_{\perp}v_{\perp}/\Omega=k_{\perp}\tan\xi/k_{\parallel,res}>1$
if the pitch angle $\xi$ is not close to 0, we can simply use
the asymptotic form of Bessel function for large argument. Then we
can derive an analytical result for Alfv\'{e}nic turbulence (see
Appendix B), \begin{equation}
\left[\begin{array}{c}
D_{\mu\mu}\\
D_{pp}\end{array}\right]=\frac{v^{2.5}\cos\xi^{5.5}}{\Omega^{1.5}L^{2.5}\sin\xi}\Gamma[6.5,k_{max}^{-\frac{2}{3}}k_{res}L^{\frac{1}{3}}]\left[\begin{array}{c}
1\\
m^{2}V_{A}^{2}\end{array}\right],\label{ana}\end{equation}
where $\Gamma[a,z]$ is the incomplete gamma function. The presence
of this gamma function in our solution makes our results orders of
magnitude larger than those%
\footnote{We compared our result with the resonant term as the nonresonant term
is spurious as noted by Chandran (2000). %
} in Chandran (2000) for the most of energies considered. However,
the scattering frequency,

\be
\nu=2D_{\mu\mu}/(1-\mu^{2}),\label{nu}
\ee
are much smaller
than the estimates for isotropic and slab model (see Yan \& Lazarian
2002). The Alfv\'{e}n modes are damped in partially ionized medium at scales
larger than the resonant scales of moderate energy CRs. 
As the anisotropy of the Alfv\'{e}n modes is increasing with the
decrease of scales, the interaction with Alfv\'{e}n modes becomes
more efficient for higher energy cosmic rays. When the Larmor radius
of the particle becomes comparable to the injection scale, which is
likely to be true in the shock region as well as for very high energy cosmic
rays in diffuse ISM, Alfv\'{e}n modes get important.

It's worthwhile to mention the imbalanced cascade of Alfv\'{e}n modes
(CLV02). Our basic assumption above was that Alfv\'{e}n modes were
balanced, meaning that the energy of modes propagating one way was
equal to that in opposite direction. In reality, many turbulence sources
are localized so that the modes leaving the sources are more energetic
than those coming toward the sources. The energy transfer in the imbalanced
cascade occurs at a slower rate, and the Alfv\'{e}n modes behave
more like waves. The scattering by these Alfv\'{e}n modes is likely
to be more efficient. However, as the degree of anisotropy of imbalanced
cascade is currently uncertain, and the process will be discussed elsewhere.

\subsection{Scattering by fast modes}

The contribution from slow modes is no more than that by Alfv\'{e}n
modes since the slow modes have the similar anisotropies and scalings.
More promising are fast modes, which are isotropic (CL02). With fast
modes there can be two types of resonant interaction: gyroresonance
and transit-time damping (TTD) (Schlickeiser \& Miller 1998).

Fast modes potentially can scatter CRs by transit-time damping. From
the resonant condition $\omega-k_{\parallel}v_{\parallel}\sim0$,
we see that the contribution is mostly from nearly perpendicular propagating
modes ($\cos\theta\sim0$). According to Eq.(4),we see that the corresponding
correlation tensor for the magnetic fluctuations $M_{ij}$ are very
small, so the contribution from TTD to pitch angle scattering is less
important unless the pitch angle is close to $90^{o}$. The advantage
of TTD is that it doesn't have a distinct resonant scale associated
with it. The resonant condition only requires $k_{\parallel}/k=V_{ph}/V_{\parallel}$.
As the result, TTD is an alternative to scatter low energy CRs whose
Larmor radii are below the damping scale of the fast modes. Moreover,
TTD can be substantial to the momentum diffusion or acceleration (also
known as the second order Fermi acceleration). It can be crucial in
some circumstances, e.g., for $\gamma$ ray burst (Lazarian et al.
2003), and super-Alfv\'{e}nic grain acceleration (Yan \& Lazarian
2003).

Here we apply our analysis to the various phases in ISM.

Recent observations (Beck 2001) suggest that the galactic halos are
magnetic-dominant, corresponding to low $\beta$ medium. For low $\beta$
medium, we use the tensors given by Eq.(4). For the halo, the collisionless
damping is dominant (see Table 1). As we see from Eq.(\ref{lbcoll}),
the damping increases with $\theta$ unless $\theta$ is close to
$\pi/2$, which results in $k_{\perp,c}<k_{\parallel,res}$. This
means that the fast modes propagating at smaller angle with the magnetic
field are less damped. As a result the argument for the Bessel function
in Eq.(\ref{genmu}) is $k_{\perp}\tan\xi/k_{\parallel,res}<1$
unless $\xi$ is close to $90^{o}$. So we can take advantage of
the anisotropy of the damped fast modes and use the asymptotic form
of Bessel function for small argument $J_{n}(x)\simeq(x/2)^{n}/n!$
to obtain the corresponding analytical result for this case (see
Appendix~B):

\begin{eqnarray}
\left[\begin{array}{c}
D_{\mu\mu}\\
D_{pp}\end{array}\right]=\frac{\pi(\Omega v\mu)^{0.5}(1-\mu^{2})}{2L^{0.5}}\nonumber \\
\left[\begin{array}{c}
(1-(\left(\frac{k_{\perp,c}}{k_{\parallel,res}}\right)^{2}+1)^{-\frac{7}{4}})/7\\
(1-(\left(\frac{k_{\perp,c}}{k_{\parallel,res}}\right)^{2}+1)^{-\frac{3}{4}})m^{2}V_{A}^{2}/3\end{array}\right]
\label{fastlb}\end{eqnarray}
From this equation, we see the key factor is $(k_{\perp,c}/k_{\parallel,res})=\tan\theta_{c}$.
As the collisionless damping is a function of $\theta$, the cutoff
scale depends also on $\theta$. Put the Eq.(\ref{lbcoll}) into the
relation $\tau_{k}\Gamma_{L}\simeq1$, we see $\cos^{1.5}\theta_{c}/\sin^{2}\theta_{c}\propto k_{\parallel,res}^{0.5}\propto\gamma^{-0.5}$.
Therefore scattering frequency in the halo increases slowly with energy
as shown in Fig.1a. For TTD, the contribution is mostly from the nearly
perpendicular propagating modes for which the collisionless damping
is small (see Eq.\ref{lbcoll}). Thus the asymptotics of the Bessel
function is not applicable. We obtain a numerical solution from Eq.(\ref{TTD}).
The contribution of TTD to scattering is much smaller than that of
gyroresonance.

Hot ionized medium (HIM) is in high $\beta$ regime and we adopt
the tensors given by Eq.(\ref{hbtensor}). The mean free path of the
HIM is of the order of parsec and thus the fast modes in this medium
are subjected to collisionless damping (see table1). We use Eq.(\ref{hbcoll})
to get the corresponding damping scale $k_{c}^{-1}=1.3\times10^{13}\sin^{4/3}\theta/\cos^2\theta$cm.
Therefore,  fast modes become also anisotropic in HIM resulting in $k_{\perp,c}<k_{\parallel,res}$. 
For CRs with energy $E_{k}\le25$GeV, $k_{\parallel,res}>k_{\perp,c}=\sqrt{k_{c}^{2}-k_{\parallel}^{2}}$,
so we can use the asymptotics of Bessel function $J_{n}(x)\simeq(x/2)^{n}/n!$
to get an analytical result for gyroresonance in high $\beta$ medium:

\begin{eqnarray}
\left[\begin{array}{c}
D_{\mu\mu}\\
D_{pp}\end{array}\right]=\frac{2\pi(\Omega v\mu)^{0.5}(1-\mu^{2})}{L^{0.5}}\nonumber\\
\left[\begin{array}{c}
(1-\left(\frac{k_{\parallel,res}}{k_{c}}\right)^{7/2})/7\beta^{2}-(1-\left(\frac{k_{\parallel,res}}{k_{c}}\right)^{11/2})/11\beta^{2}\\
(1-\left(\frac{k_{\parallel,res}}{k_{c}}\right)^{3/2})m^{2}V_{A}^{2}/3\beta-(1-\left(\frac{k_{\parallel,res}}{k_{c}}\right)^{7/2})m^{2}V_{A}^{2}/7\beta\end{array}\right],\label{fasthb}\end{eqnarray}
from which we see the scattering frequency decreases with $k_{\parallel,res}$
and so increases with $E_{k}$. This result agrees well with the numerical
solution obtained by Eq.(\ref{gyro}) (see Fig.1a).
At high energy end, the resonant scales are much larger than the damping
scale $k_{c}^{-1}$ so the effect of damping to the scattering curve
can be ignored. As a result, the shape of the scattering curve is
similar to the case where there is no damping (see Yan \& Lazarian
2002). The scattering frequency in this case according to Eq.(\ref{gyro}),
increases with Larmor frequency $\Omega_{0}/\gamma$ and the resonant
scale $k_{res}^{-1}\sim r_{L}$ , which increase with proton energy
$E_{k}$. For $E_{k}>1GeV$, $\gamma$ goes nearly linearly with $E_{k}$
so that the scattering frequency decreases due to the $\Omega_{0}/\gamma$
dependence which changes more rapidly.

According to the results in $\S$2, the warm ionized medium (WIM)
is also in low $\beta$ but collisional regime. The mean free path $l_{mfp}=6\times10^{12}$cm
(table1), which corresponds to the resonant scale of CRs with energy
$\sim10$GeV. Thus For CRs harder than 10GeV the resonant modes with $k_{\parallel}=\Omega/v_\parallel$  are subjected to collisional damping. Since the viscous damping increases with $\theta$, we can
apply Eq.(\ref{fastlb}) to the gyroresonance in WIM as well. As
the viscous damping in the low $\beta$ medium scales as $k_{\perp}^{2}$while
the cascading rate $\tau_{k}^{-1}\propto k^{0.5}$, we can get $\cos^{1.5}\theta_{c}/\sin^{2}\theta_{c}\propto k_{\parallel,res}^{1.5}\propto\gamma^{-1.5}$ from
the truncation condition $\tau_{k}\Gamma_{L}\simeq1$. Thus according to Eq.(\ref{fastlb})
the scattering frequency curve in WIM gets steeper (see Fig.1a). For
lower energy CRs, the only available modes are those with ${\bf k}$ in a small cone as the residual of the collisional damping at large scales. As a conservative estimate, we adopt the smallest perpendicular viscous cutoff scale $k_{\perp,c}$ above the mean free path to do the calculation. Certainly, the resonant modes corresponding to this energy range of CRs  are also affected by collisionless damping. However, the pitch angle of these modes are so small that the collisionless process is marginal (see Eq.(\ref{lbcoll})).

In partially ionized medium, the fast modes are severely damped by
the ion-neutral damping. The cascade is cut off before the resonant
scales for most of the energy range we considered. Indeed, for the parameters chosen,
we find that gyroresonance only contribute to the transport of CRs
of energy $\geq0.8$TeV in DC and $>1$TeV for WNM and CNM. Therefore
only TTD contributes to the scattering of moderate energy CRs in these
media. Since $k_{\perp}v_{\perp}/\Omega\le k_{c}\tan\xi/k_{\parallel,res}<1$,
we can use the asymptotics of Bessel function $J_{n}(x)\simeq(x/2)^{n}/n!$
to get an analytical result for TTD in these media (see Appendix~B)

\begin{equation}
\left[\begin{array}{c}
D_{\mu\mu}\\
D_{pp}\end{array}\right]=\frac{\pi(1-\mu^{4})^{2}k_{c}^{0.5}V_{A}^{2}}{L^{0.5}v\mu^{3}}\left(1-\left(\frac{V_{A}}{v\mu}\right)^{2}\right)^{1.5}\left[\begin{array}{c}
1\\
(p\mu)^{2}\end{array}\right],\label{lbttd}\end{equation}
from which we see $D_{\mu\mu}$ is approximately $\propto1/v$. As
$v$ increases with $E_{k}$ when $E_{k}<m_{p}c^{2}\simeq1$GeV and
$\simeq c$ when $E_{k}>1$GeV, the scattering frequency decreases
with energy when $E_{k}<1$GeV and keeps nearly a constant at higher
energies. As expected, the
scattering is much less efficient in partially ionized media because
of ion-neutral damping.

\begin{figure}
\leavevmode
\includegraphics[ width=0.45\columnwidth]{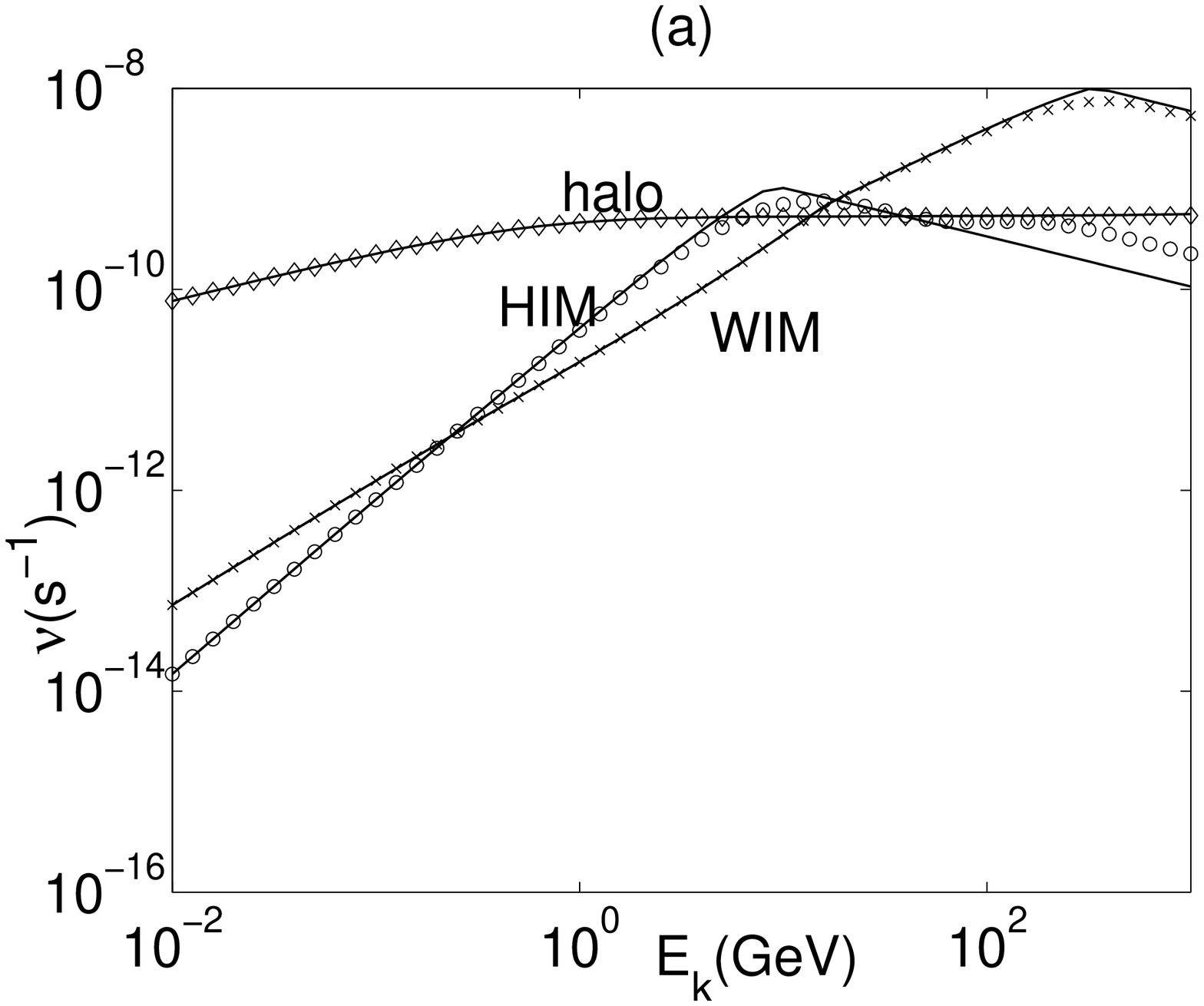} \hfil
\includegraphics[  width=0.45\columnwidth]{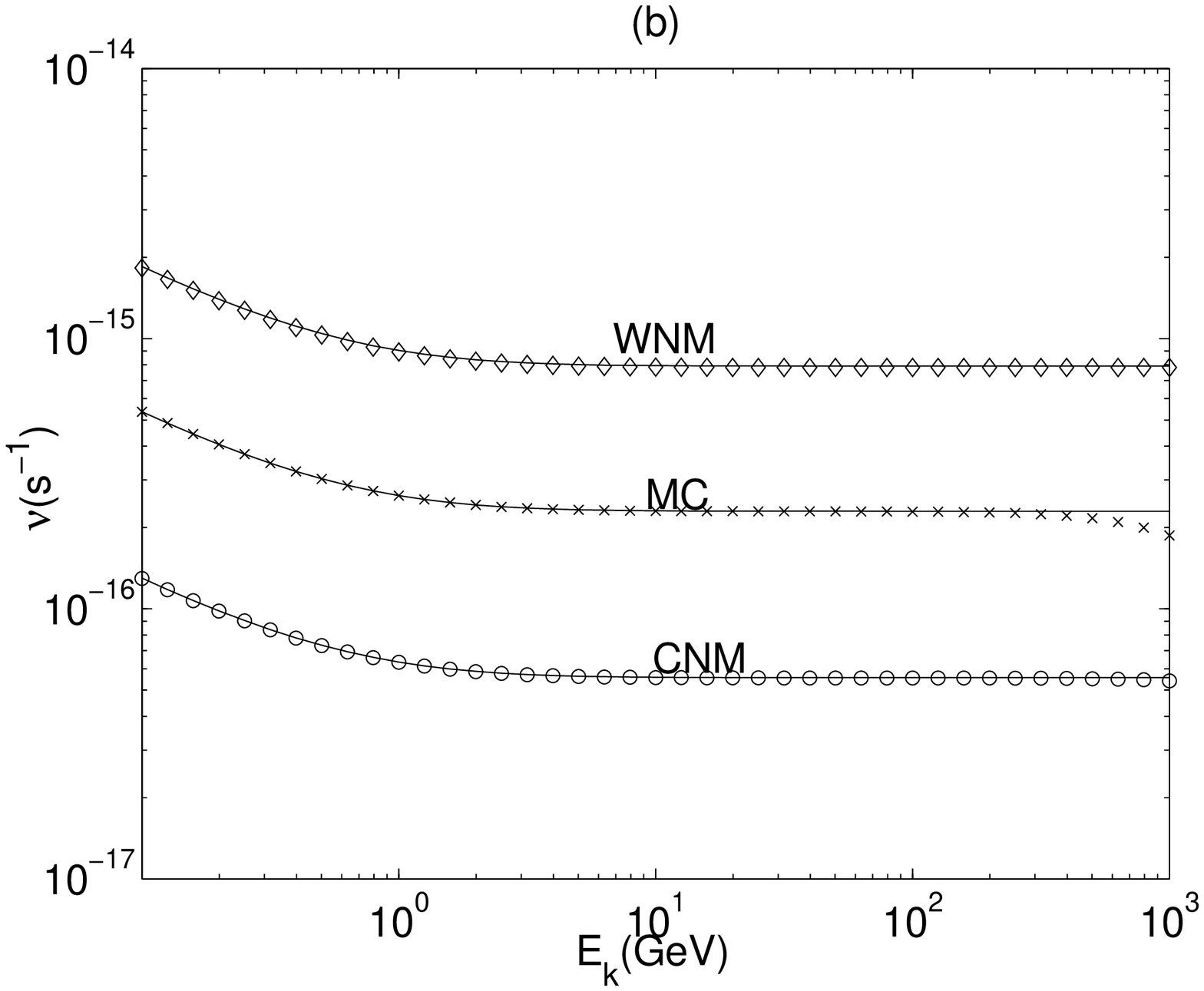}
\caption{Scattering by fast modes. The scattering frequency $\nu$ given by Eq.(\ref{nu}) vs. (a) the kinetic energy $E_{k}$ of
cosmic rays,  (b) kinetic energy $E_{k}$ of
cosmic rays in partially ionized
medium. The solid lines represent our analytical results. In (a),
the diamond line is our numerical result for galactic halo, the 'o' line refers
to HIM and the 'x' line represents WIM. In (b), the diamond
line is our numerical result for WNM, the 'o' line refers to CNM and
the 'x' line represents MC.}
\end{figure}

From the above results, we see that the fast modes dominate
CRs' scattering in spite of the damping. From Eq.(\ref{lbttd}), we see the ratio of scattering and momentum diffusion
rates $p^{2}D_{\mu\mu}/D_{pp}=1/\mu^{2}$, which means
that the scattering and acceleration by TTD are comparable. However,
for gyroresonance, we know that $p^{2}D_{\mu\mu}/D_{pp}\sim v^{2}/(min\{\beta,1\}V_{A}^{2})$
(see Eq.(\ref{fasthb})and Eq.(\ref{fastlb})). Consequently for momentum
diffusion (or acceleration) TTD is dominant and the rate differs from
the scattering rate by a factor $\sim \mu^{2}$. 

A special case is that the cosmic rays propagate nearly perpendicular
to the magnetic field, so called the $90^{o}$ scattering problem.
The dynamics of the turbulence was suggested to be taken into account in this case (Schlickeiser \& Achatz
1993). There have been attempts of using the concept of the dynamical
turbulence in dealing with the CR transport before (Bieber et al. 1994). It has been shown in Schlickeiser \& Achatz (1993) that if the decay time $\alpha kV_A$ is longer, the resonance function can be approximated by $\delta$ function. In other words, $\alpha\rightarrow 0$ is the "magnetostatic limit". From (Eq.(\ref{fdecay})),  we can get $\alpha=(kL)^{-1/2}$. For moderate energy CRs, the corresponding $\alpha<<1$ so that $\delta$ function is still a good approximation even for the $90^o$ case. The corresponding resonant scale is $\sim V_{ph}/\Omega$. Thus unlike with Alfv\'en modes, we can implement unified $\delta$ function and then integrate over the pitch angle $\mu$ to estimate the spatial diffusion. Here we focus on galactic halo since the scattering rate is the highest there%
\footnote{It looks from Fig.1a that the scattering is most efficient in WIM for CR$>$10GeV. However, it's just for a particular $\mu$. For small $\mu$, which corresponds to small scales and is the most important for the calculation of mean free path, the scattering in halo is still dominant.}. In the precedent calculations, we ignored $V_{ph}$ because it is negligible comparing with $v\mu$. Now $v\mu$ should be replaced with $v\mu\pm V_{ph}$ as seen from the gyroresonance condition $\omega-k_{\parallel}v_\parallel=n\Omega$. The complication comes from the cutoff due to damping. As addressed earlier, the fast modes become slab-type at small scales. Thus we can get $\theta_c$ from Eq.(\ref{fdecay}) and Eq.(A4) $k_{\perp,c}/k_{\parallel,res}\simeq \theta_c\simeq 2(m_i v_\parallel/\pi m_e\beta \Omega l)^{1/4}\ll 1$. Then we can obtain an analytical solution from Eq.(\ref{fastlb}) for the parallel spatial diffusion coefficient:

\begin{eqnarray}
\kappa_\parallel&=&\frac{v^2}{8}\int_{-1}^1d\mu\frac{(1-\mu^2)^2}{D_{\mu\mu}}\nonumber \\
&=&\frac{3lv}{4}\left(\frac{m_e\beta}{m_i\pi}\right)^{0.5}[-1+2\epsilon+2(-1+\epsilon^2)\ln\frac{\epsilon}{(1+\epsilon)}],
\end{eqnarray}
where $\epsilon=V_A/v$.  This result is compared with numerical integral of Eq.(\ref{fastlb}) in Fig.2b. These results are not applicable to high energy CRs for which the slab-model of fast modes fails. 

\begin{figure}
\includegraphics[ width=0.45\columnwidth]{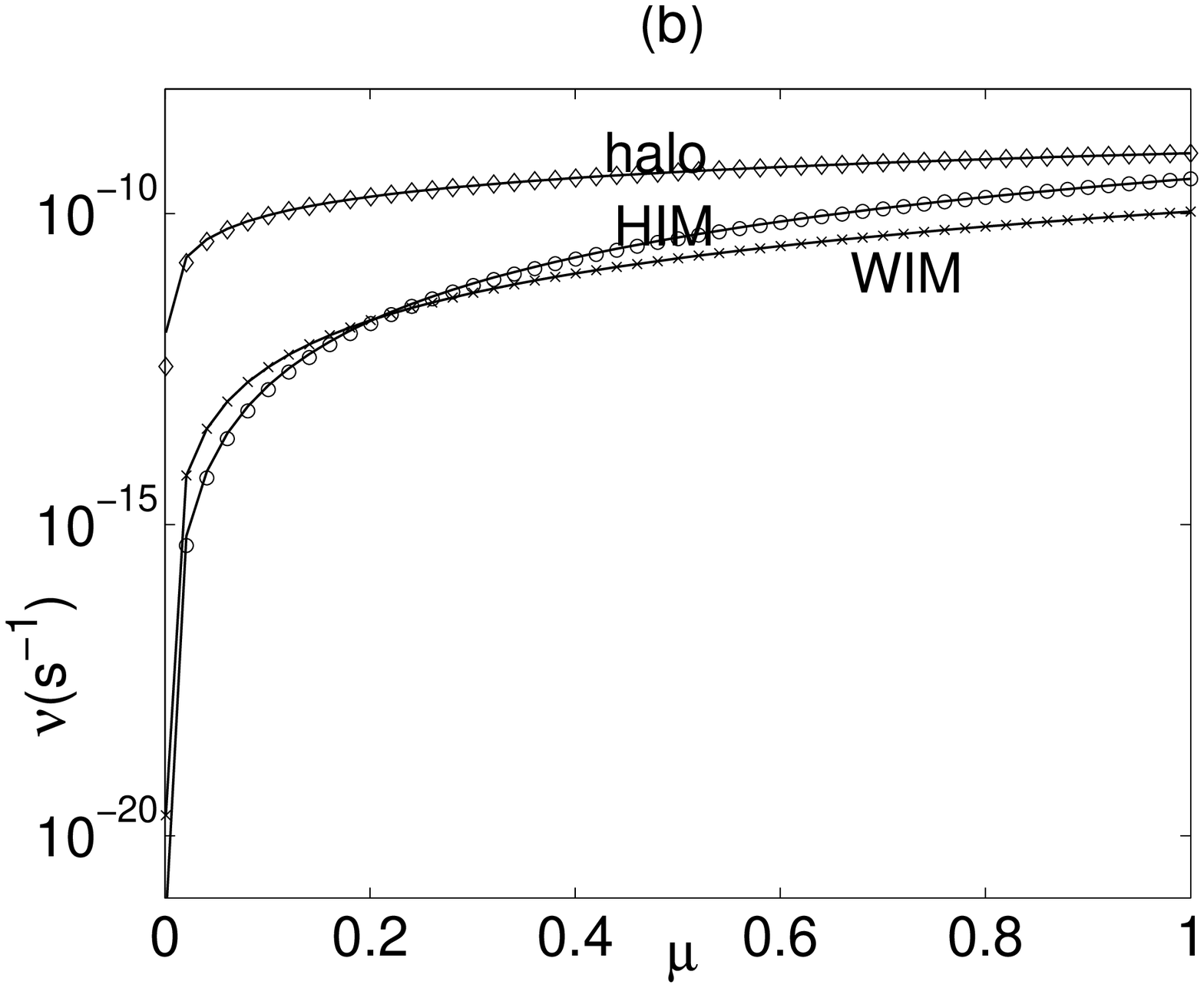} \hfil 
\includegraphics[ width=0.45\columnwidth]{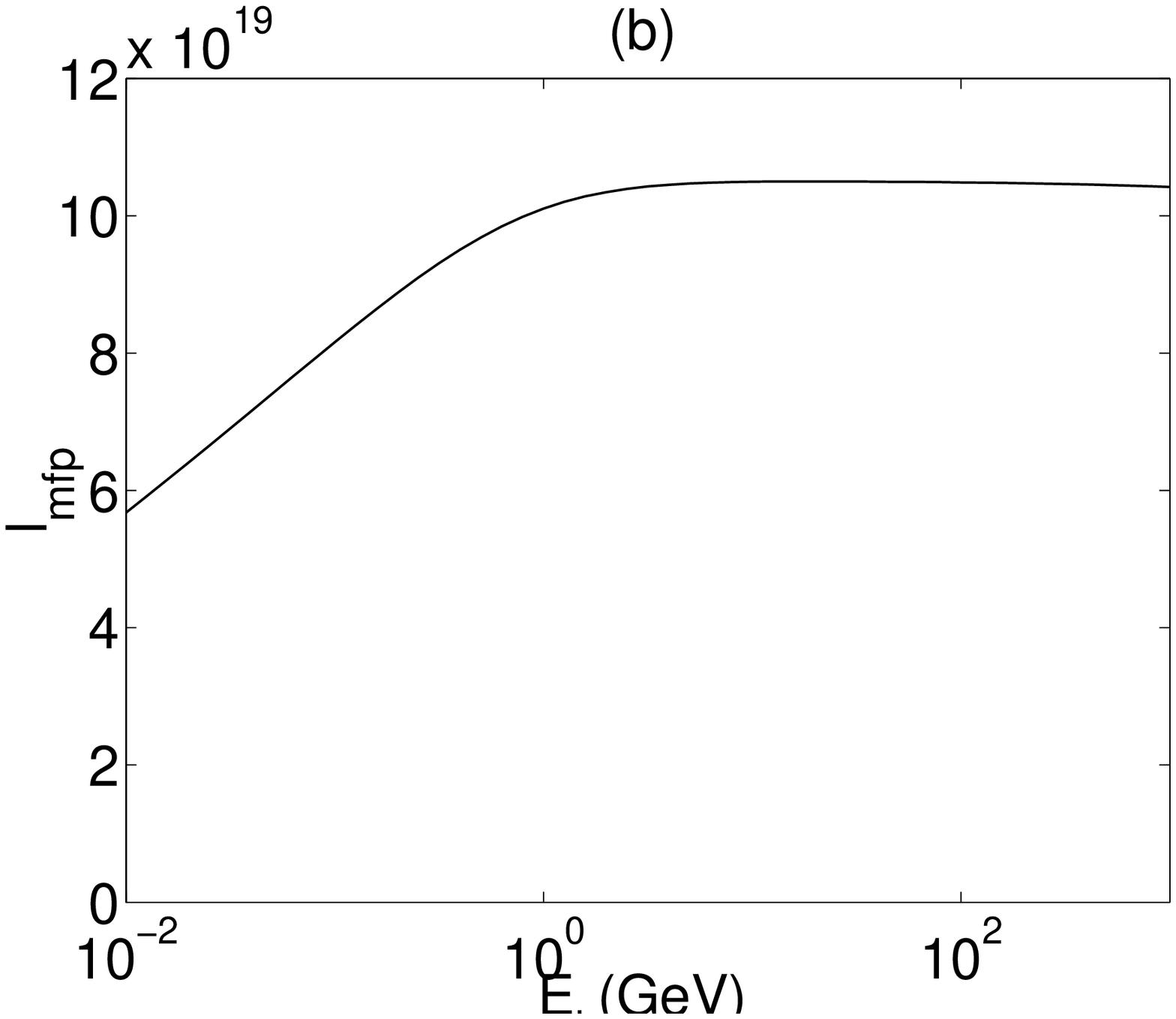}
\caption{(a)The scattering frequency $\nu$ vs. cosine of pitch angle $\mu$ of 1GeV CR in fully ionized medium. The solid lines represent our analytical results, the diamond line is our numerical result for galactic halo, the 'o' line refers
to HIM and the 'x' line represents WIM. (b), The mean free path of CRs in galactic halo resulting from numerical integration of Eq.(\ref{fastlb})}.
\end{figure}

It should be pointed out that all the results above are very much
dependent on the plasma $\beta$, which is somewhat uncertain. But
the basic conclusion is definite: fast modes dominate the cosmic ray
scattering except in very high $\beta$ medium where fast modes are
severely damped. This demands a substantial revision of cosmic ray
acceleration/propagation theories, and many related problems should be revisited.

\section{Cosmic ray self confinement by streaming instability}

When cosmic rays stream at a velocity much larger than Alfv\'{e}n
velocity, they can excite resonant MHD modes which in turn scatter
themselves. This is so called streaming instability. It was claimed
that the instability could provide confinement for cosmic rays with
energy less than 100GeV (Cesarsky 1980). However, this was calculated
in an ideal regime, namely, there was no background MHD turbulence.
Only ambipolar collisional damping was involved. In other words, the
self-excited modes would not be damped in fully ionized gas, which
is not true for turbulent medium. Here we shall reconsider in turbulent
ISM how this instability would affect cosmic ray transport, whether
it is still an effective mechanism or not.

The growth rate of the modes of wave number $k$ is given by (Longair
1997)

\begin{mathletters}
\begin{equation}
\Gamma(k)=\Omega_0\frac{N(\geq E)}{n_{p}}(-1+\frac{v}{V_{A}}),\label{instability}\end{equation}
 where $N(\geq E)$ is the number density of cosmic rays with energy
$\geq E$ which resonate with the wave, $i.e.$, $v(E)=\Omega/k_{\parallel}$,
$n_{p}$ is the number density of charged particles in the medium.
If taking the energy spectrum of cosmic ray particles $n(E)\sim1.6\times10^{5}(E/$MeV$)^{-2.7}$
as measured at the Earth (Longair 1997), then $N(\geq E)\simeq0.25\times10^{-10}(E/$GeV$)^{-1.7}$cm$^{-3}$sr$^{-1}$.

The damping rate due to the turbulence cascade depends on the properties of turbulence. To estimate the rate at which background turbulence can suppress th streaming instability we consider both interaction with fast and Alfv\'enic modes. The interaction with the fast modes happens at the rate $\tau_k\sim (k/L)^{-1/2}V_{ph}/V^2$ (see Eq.(\ref{fdecay})). The interaction with Alfv\'en modes is less straightforward to quantify because of their anisotropy.
Consider the interaction of a perturbation created by streaming particles and the background Alfv\'enic turbulence. If we approximate this perturbation by a wave with $k=k_\parallel$, it gets evident that the interaction will be very different from what takes place between the oppositely moving wave packets in Alfv\'enic turbulence. If we consider local in ${\bf k}$-space interaction the plane wave created by the streaming instability will undergo random shearing. The mean variation of the phase $\Delta \varphi$ will be approximately $1/\sqrt N$, where $N\sim (k_\perp/k_\parallel)^2$. The decorrelation time $t_{dc}$ can be estimated using random walk arguments $(\Delta \varphi)^2\cdot (t_{dc}\omega)\sim 1$, which provides $t_{dc}\sim \omega^{-1} (k_\perp/k_\parallel)^2$.

Thus the dominant damping mechanism here is the cascade of fast modes. By equating the growth rate Eq.((\ref{instability})) and the damping rate (Eq.(\ref{fdecay})), we can find that the streaming instability
is only applicable for particles with energy $<1.5\times 10^{-9}[n_{p}^{-1}(v/V_{A})^{1.5}(L\Omega_0/V_A)^{0.5}]^{1/1.2}$GeV.
In HIM, the corresponding energy is $\sim 500$GeV. As shown in the Eq.(\ref{instability}) , the
growth rate depends on the CRs' density. In those regions where high
energy particles are produced, e.g., shock fronts in ISM, $\gamma$ ray burst, SN, the
streaming instability is more important.

\end{mathletters}

\section{Discussion}

We have attempted to apply a realistic model recently obtained for
MHD turbulence to the problem of cosmic ray scattering taking into
account the turbulence cutoff arising from both collisional and collisionless
damping. We considered both gyroresonance
and transit-time damping (TTD),  and used QLT to obtain our estimates.

In earlier papers it was frequently 
assumed that Alfv\'enic turbulence follows Kolmogorov 
spectrum while
the problem of CR scattering was considered. This is incorrect because
MHD turbulence is different from hydrodynamic turbulence if magnetic
field is dynamically important. Realistic
MHD turbulence can be decomposed into, Alfv\'{e}n, slow
and fast modes. They have different statistics (see CL02). The Alfv\'{e}n
and slow modes follow GS95 scalings and show scale-dependent anisotropy.
The anisotropy of Alfv\'{e}n and slow modes reduce the scattering
efficiency to such a degree that their impact on cosmic ray propagation
is marginal. 

Fast modes, however, are promising as a means of scattering due to their 
isotropy. Even in spite of the damping fast modes dominate scattering if 
turbulent
energy is injected at large scales. We provided calculations for various
phases of ISM, including galactic halo, HIM, WIM and partially ionized media. As the fast
modes are subjected to different damping processes, the CR 
scattering varies . In galactic halo and HIM, the fast
modes are subjected to collisionless damping.
The mean free path in WIM is
much smaller, and ion viscosity dominates the damping. All these dampings 
decrease with $\theta$ and marginal for parallel modes. As the result, the 
fast modes at small scales become anisotropic with 
$k_{\parallel}\gg k_{\perp}$. In this sense, fast modes are in slab geometry 
but with less energies. These modes can efficiently scatter CRs with the 
resonant scale $k_{\parallel,res}=\Omega/v_{\parallel}$
larger than the cutoff through gyroresonance. TTD is usually weaker
for pitch angle scattering than gyroresonance, but dominant for momentum
diffusion. In all the cases, the scattering efficiency decreases with plasma 
$\beta$
because the damping increases with $\beta$ and the perturbations
of turbulence decreases with $\beta$.

 How good is our model of ISM turbulence? We used spectra for Alfv\'en and 
fast modes which follow from theoretical constructions and which were tested 
numerically (CL02). Within the ISM similar spectra develop 
dynamically when random driving is applied at large scales (see Mac Low 2003) 
and therefore we expect that the distribution of energy over scales for 
different modes to coincide with the assumed ones. The partition of energy 
over modes depend, however, on driving. For random driving with 
$\delta V\sim V_A$ we expect approximately equal energy in fast and Alfv\'{e}n 
modes. As the details of ISM driving are not clear and the QLT provides 
results correct up to a factor of the order unity, our predictions for fast 
mode scattering
contain an uncertain factor of the order unity. We believe, however, 
that our results for scattering by Alfv\'en modes that given by Eq.(\ref{ana}) 
are subject to more uncertainties related to the turbulence driving. 
Indeed, the scattering rates that follow from 
Eq.(\ref{ana}), although orders of magnitude larger than those in Chandran 
(2000), are still {\it extremely} small compared to models with 
isotropic spectra. Because it is so small, higher order effects may become
important.
Does the steady state anisotropic spectrum has always enough time to develop?
The dynamic establishing of anisotropic spectra of Alfv\'{e}nic turbulence 
requires a few eddy turnover times. For imbalanced driving this can
take somewhat longer time (see discussion in $\S$5). Therefore
one can imagine situations when the spectrum deviates of its
steady state form given by Eq. (3). These deviations are most
likely to be small. Therefore we do not expect that in typical ISM
conditions the contribution of Alfv\'{e}nic
modes is comparable to that of the fast modes. Identification of
the particular situations when the opposite may be true requires
a more systematic study of particular driving mechanisms and is
the subject of future research. We also note, that if energy is injected
at the small scales, the anisotropy of the Alfv\'{e}nic modes is not
large and the scattering can be appreciable. The
deviations of the spectrum from the steady state one are not important.

Although scattering via fast modes is isotropic at large scales and becomes 
slab-type at small scales at which damping gets important,  our model is 
radically different from earlier discussed slab and isotropic+slab models 
(Bieber et al. 1994, Schlickeiser \& Miller 1998).We used theoretically 
motivated models of turbulence that have been tested numerically. Therefore 
the transitions from one regime to another as well as the values of the 
intensity of perturbations have good justification within our approach.

We adopted a particular set of parameters of ISM while doing the calculation.
The basic conclusion, namely, fast modes dominate the cosmic ray scattering,
will remain the same if the parameters are altered. However, since
the $\beta$ value in ISM remains uncertain from observation,
the damping scale $k_{c}$ is somewhat uncertain. And this determines
the energy limit down to which gyroresonance is dominant.

\section{Summary}

In the paper above we characterized interaction of cosmic rays with balanced interstellar turbulence driven at a large scale. Our results can be summarized as follows:

1. Fast modes provide the dominant contribution to cosmic ray scattering.

2. Scattering of cosmic rays varies from one interstellar phase to another. It sensitively depends on the damping of fast modes.

3. Due to anisotropic plasma damping fast modes develop at small scales anisotropy which makes gyroresonant scattering within slab approximation applicable. At scales where damping is negligible the isotropic gyroresonant scattering approximation is applicable.

4. In partially ionized gas fast modes scatter cosmic rays of energies $\leq 1$TeV by TTD mechanism.

5. Streaming instability is partially suppressed due to the interaction of the emerging magnetic perturbations with the surrounding turbulence.

\begin{acknowledgments}
We acknowledge valuable discussions with Jungyeon Cho, Steve Cowley, Don Cox, Randy Jokipii, Vladimir Mirnov, and Reinhard Schlickeiser. This work is supported by the NSF grant AST0125544.

\end{acknowledgments}

\appendix

\section{A. Damping of MHD turbulence}
Below we summarize the damping processes that we consider in the paper.

\emph{Ion-neutral damping}

In partially ionized medium, viscosity of neutrals provides damping
(see LY02). If the mean free path for a neutral atom, $l_{n}$, in
a partially ionized gas with density $n_{tot}=n_{n}+n_{i}$ is much
less than the size of the eddies under consideration, i.e. $l_{n}k\ll1$,
the damping time

\begin{equation}
\Gamma_{in}\sim\nu_{n}^{-1}k^{-2}\sim\left(\frac{n_{tot}}{n_{n}}\right)(l_{n}c_{n})^{-1}k^{-2},\label{tdamp}\end{equation}
 where $\nu_{n}$ is effective viscosity produced by neutrals%
\footnote{The viscosity due to ion-ion collisions is typically small as ion
motions are constrained by the magnetic field. %
}. The mean free path of a neutral atom $l_{n}$ is influenced both
by collisions with neutrals and with ions. The rate at which neutrals
collide with ions is proportional to the density of ions, while the
rate at which neutrals collide with other neutrals is proportional
to the density of neutrals. The drag coefficient for neutral-neutral
collisions is $\sim1.7\times10^{-10}T$(K)$^{0.3}$ cm$^{3}$ s$^{-1}$
(Spitzer 1978), while for neutral-ion collisions it is $\sim{\langle v_{r}\sigma_{in}\rangle}\approx1.9\times10^{-9}$
cm$^{3}$ s$^{-1}$ (Draine, Roberge \& Dalgarno 1983). Thus collisions
with other neutrals dominate for $n_{i}/n_{n}$ less than $\sim0.09T^{0.3}$.
Turbulent motions cascade down till the cascading time is of the order
of $t_{damp}$. The maximal damping corresponds to $k^{-1}\sim l_{n}$.
If the neutrals constitute less than approximately $5\%$, the cascade
goes below $l_{n}$ and is damped at smaller scale (see below).

\emph{Collisionless damping}

The nature of collisionless damping is closely related to the radiation
of charged particles in magnetic field. Since the charged particles
can emit plasma modes through acceleration (cyclotron radiation) and
Cherenkov effect, they also absorb the radiation under the same condition
(Ginzburg 1961). The thermal particles can be accelerated either by
the parallel electric field which can also be called Landau damping
or the magnetic mirror (TTD) associated with the comoving modes (or
under the Cherenkov condition $k_{\parallel}v_{\parallel}=\omega$).
The gyroresonance with thermal ions also causes the damping of those
modes with frequency close to the ion-cyclotron frequency (Leamon et al. 1998),  though
it is irrelevant to the low frequency modes since we deal with GeV CRs here. The
collisionless damping depends on the plasma $\beta\equiv P_{gas}/P_{mag}$
and the propagation direction of the modes. In general, the damping
increases with the plasma $\beta$. And the damping is much more severe
for fast modes than for Alfv\'{e}n modes. For instance, the Alfv\'{e}n
modes are weakly damped even in a high $\beta$ medium, where fast
modes are strongly damped. The damping rate $\gamma_{d}=\tau_{d}^{-1}$
of the fast modes of frequency $\omega$ for $\beta\ll1$ and $\theta\sim1$
(Ginzburg 1961) is

\begin{eqnarray}
\Gamma_{L} & = & \frac{\sqrt{\pi\beta}}{4}\omega\frac{\sin^{2}\theta}{\cos\theta}\times[\sqrt{\frac{m_{e}}{m_{H}}}\exp(-\frac{m_{e}}{m_{H}\beta\cos^{2}\theta})\label{lbcoll}\\
 & + & 5\exp(-\frac{1}{\beta\cos^{2}\theta})],\end{eqnarray}
 where $m_{e}$ is the electron mass. The exact expression for the
damping of fast modes at small $\theta$ was obtained in Stepanov%
\footnote{We corrected a typo in the corresponding expression.%
} (1958)

\[
\Gamma_{L}=\frac{\sqrt{\pi\beta}}{4}\omega\theta^{2}\times\left(1+\frac{\theta^{2}}{\sqrt{\theta^{4}+4\Omega_{i}^{2}/\omega^{2}}}\right)\sqrt{\frac{m_{e}}{m_{H}}}\exp(-\frac{m_{e}}{m_{H}\beta\cos^{2}\theta}).\]
 When $\beta\gg1$, we obtain the damping rate as a function of $\theta$ from Foote \& Kulsrud 1979,

\begin{eqnarray}
\Gamma_{L}=\frac{\sin^2\theta}{\cos^3\theta}\times \cases{2\omega^{2}/\Omega_{i}  &  for $k<\Omega_{i}/\beta V_{A}$\cr
2\Omega_{i}/\beta  &  for $k>\Omega_{i}/\beta V_{A}$\cr}
\label{hbcoll}
\end{eqnarray}
 where $\Omega_{i}$ is the ion gyrofrequency.

\emph{Ion viscosity}

In a strong magnetic field ($\Omega_{i}\tau_{i}\gg1$) the transport
of transverse momentum is prohibited by the magnetic field. Thus transverse
viscosity $\eta_{\perp}$ is much smaller than longitudinal viscosity
$\eta_{0}=0.96n(T/eV)\tau_{i}$, $\eta_{\perp}\sim\eta_{0}/(\Omega_{i}\tau_{i})^{2}$.
The heat generated by this damping is $Q_{vis}=\eta_{0}(\partial v_{x}/\partial x+\partial v_{y}/\partial y-2\partial v_{z}/\partial z)^{2}/3$,
where $v_{x}$, $v_{y}$, $v_{z}$ are the velocity components of
the wave perturbation (Braginskii 1965). From the expression, we see
that the viscous damping is not important unless there is compression.
Therefore Alfv\'{e}n modes is marginally affected by the ion viscosity.

While the damping due to compression along the magnetic fields can
be easily understood, it is counterintuitive that the compression
perpendicular to the magnetic also results in damping through longitudinal
viscosity. Following Braginskii (1965), the damping of perpendicular
motion may be illustrated in the following way. To understand the
physics of such a damping, consider fast modes in low $\beta$ medium.
In this case, the motions are primarily perpendicular to the magnetic
field so that $\partial v_{x}/\partial x=\dot{n}/n\sim\dot{B}/B$.
The transverse energy of the ions increases because of the conservation
of adiabatic invariant $v_{\perp}^{2}/B$. If the rate of compression
is faster than that of collisions, the ion distribution in the momentum
space is bound to be distorted from the Maxwellian isotropic sphere
to an oblate spheroid with the long axis perpendicular to the magnetic
field. As a result, the transverse pressure gets greater than the
longitudinal pressure by a factor $\tau_{i}\sim\dot{n}/n$, resulting
in a stress $\sim p\tau_{i}\dot{n}/n\sim\eta_{0}\partial v_{x}/\partial x$.
The restoration of the equilibrium increases the entropy and causes
the dissipation of energy with a damping rate $\Gamma_{ion}=k_{\perp}^{2}\eta_{0}/6\rho_{i}$
(Braginskii 1965).

In high $\beta$ medium, the velocity perturbations are radial as
pointed out in $\S2$. Thus $Q_{vis}=\eta_{0}(k_{x}v_{x}+k_{y}v_{y}-2k_{z}v_{z})^{2}/3=\eta_{0}k^{2}v^{2}(\sin^{2}\theta-2\cos^{2}\theta)^{2}/3$.
Dividing this by the total energy associated with the fast modes $E_{k}=\rho_{i}v^{2}$,
we can obtain the damping rate $\Gamma_{ion}=k^{2}\eta_{0}(1-3\cos^{2}\theta)^{2}/6\rho_{i}$.

All in all,

\begin{eqnarray}
\Gamma_{ion}=\cases{	k_{\perp}^{2}\eta_{0}/6\rho_{i}   & for $\beta\ll 1$\cr
					k^{2}\eta_{0}(1-3\cos^{2}\theta)^{2}/6\rho_{i} &  for $\beta\gg 1$\cr}
\label{viscous}
\end{eqnarray}

\emph{Resistive damping}

In the paper we ignore the resistive damping because of the following reason. The resistivity is much
smaller than the longitudinal ion viscosity for fast modes. For parallel
propagating fast modes which are not subjected to the viscous damping
from the longitudinal viscosity, the resistive damping scale can be
obtained by equating the damping rate $c^{2}k^{2}/8\pi\sigma_{\perp}$
with the cascading rate of the fast modes, where $\sigma_{\perp}\simeq10^{13}(T/eV)^{3/2}$
(see Kulsrud \& Pearce 1969) is the conductivity perpendicular to
the magnetic field: $k_{res}^{-1}=1.8\times10^{7}(L/pc)^{1/3}(T/10^{4}K)^{-1}(V_{A}/kms^{-1})^{-2/3}$cm.
This scale turns out to be much smaller than the mean free path, where
collisionless damping takes over. For Alfv\'{e}n modes, the resistive
damping rate is $\Gamma_{res}=c^{2}k_{\perp}^{2}/8\pi\sigma_{\parallel}$
(the term associated with $k_{\parallel}$is negligible because of
anisotropy), where $\sigma_{\parallel}=1.96\sigma_{\perp}$is the
conductivity parallel to the magnetic field. By equating it with the
cascading rate of Alfv\'{e}n modes, we can get the resistive damping
scale $k_{res}^{-1}\simeq L(c^{2}/8\pi\sigma_{\parallel}V_{A}L)^{3/4}$.
Comparing with the proton Larmor radius, we find $k_{res}^{-1}/r_{L}=3.6\times10^{-3}(L/pc)^{1/4}(T/10^{4}K)^{-3/2}n^{1/2}(V_{A}/C_{s})^{1/4}$.
It's clear that in many astrophysical plasmas the resistive scale
is less than proton Larmor radius, where Alfv\'{e}n modes can not
proceed further because of anomalous resistivity. All in all, we see
that the damping arising from resistivity is marginal.

\section{B. Fokker-Planck coefficients}

In quasi-linear theory (QLT), the effect of MHD modes is studied by
calculating the first order corrections to the particle orbit in the
uniform magnetic field, and the ensemble-averaging over the statistical
properties of the MHD modes (Jokipii 1966, Schlickeiser \& Miller
1998). Obtained by applying the QLT to the collisionless Boltzmann-Vlasov
equation, the Fokker-Planck equation is generally used to describe
the evolvement of the gyrophase-averaged particle distribution. The
Fokker-Planck coefficients $D_{\mu\mu},D_{\mu p},D_{pp}$ are the
fundamental physical parameter for measuring the stochastic interactions,
which are determined by the electromagnetic fluctuations:

$\begin{array}{cc}
<B_{\alpha}(\mathbf{k})B_{\beta}^{*}(\mathbf{k'})>=\delta(\mathbf{k}-\mathbf{k'})P_{\alpha\beta}(\mathbf{k}) & <B_{\alpha}(\mathbf{k})E_{\beta}^{*}(\mathbf{k'})>=\delta(\mathbf{k}-\mathbf{k'})T_{\alpha\beta}(\mathbf{k})\\
<E_{\alpha}(\mathbf{k})B_{\beta}^{*}(\mathbf{k'})>=\delta(\mathbf{k}-\mathbf{k'})Q_{\alpha\beta}(\mathbf{k}) & <E_{\alpha}(\mathbf{k})E_{\beta}^{*}(\mathbf{k'})>=\delta(\mathbf{k}-\mathbf{k'})R_{\alpha\beta}(\mathbf{k}),\end{array}$

Adopting the approach in Schlickeiser \& Achatz (1993), we can get
the Fokker-Planck coefficients,

\begin{eqnarray}
\left[\begin{array}{c}
D_{\mu\mu}\\
D_{\mu p}\\
D_{pp}\end{array}\right]&=&{\frac{\Omega^{2}(1-\mu^{2})}{2B_{0}^{2}}}{\mathcal{R}}e\sum_{n=-\infty}^{n=\infty}\int_{\bf k_{min}}^{\bf k_{max}}dk^{3}  \left[i\begin{array}{c}
\left(1+\frac{\mu V_{A}}{v\zeta}\right)^{2}\\
\left(1+\frac{\mu V_{A}}{v\zeta}\right)mc\\
m^{2}c^{2}\end{array}\right]\int_{0}^{\infty}dte^{-i(k_{\parallel}v_{\parallel}-\omega+n\Omega)t}\left\{ J_{n+1}^{2}({\frac{k_{\perp}v_{\perp}}{\Omega}})\left[\begin{array}{c}
P_{{\mathcal{RR}}}({\mathbf{k}})\\
T_{{\mathcal{RR}}}({\mathbf{k}})\\
R_{{\mathcal{RR}}}({\mathbf{k}})\end{array}\right]\right.\nonumber \\
&+&J_{n-1}^{2}({\frac{k_{\perp}v_{\perp}}{\Omega}})\left[\begin{array}{c}
P_{{\mathcal{LL}}}({\mathbf{k}})\\
-T_{{\mathcal{LL}}}({\mathbf{k}})\\
R_{{\mathcal{LL}}}({\mathbf{k}})\end{array}\right]+J_{n+1}({\frac{k_{\perp}v_{\perp}}{\Omega}})J_{n-1}({\frac{k_{\perp}v_{\perp}}{\Omega}})\left.\left[e^{i2\phi}\left[\begin{array}{c}
-P_{{\mathcal{RL}}}({\mathbf{k}})\\
T_{{\mathcal{RL}}}({\mathbf{k}})\\
R_{{\mathcal{RL}}}({\mathbf{k}})\end{array}\right]+e^{-i2\phi}\left[\begin{array}{c}
-P_{{\mathcal{LR}}}({\mathbf{k}})\\
-T_{{\mathcal{LR}}}({\mathbf{k}})\\
R_{{\mathcal{LR}}}({\mathbf{k}})\end{array}\right]\right]\right\} \label{genmu}\end{eqnarray}
 where $|{\bf k_{min}}|=k_{min}=L^{-1}$, $|{\bf k_{max}}|=k_{max}$ corresponds to the dissipation
scale, where $\zeta=1$ for Alfv\'{e}n modes, $dk^3=2\pi k_\perp dk_\perp dk_\parallel$, and $\zeta=k_{\parallel}/k$
for fast modes, $m=\gamma m_{H}$ is the relativistic mass of the
proton, $v_{\perp}$ is the particle's velocity component perpendicular
to $\mathbf{B}_{0}$, $\phi=\arctan(k_{y}/k_{x}),$ ${\mathcal{L}},{\mathcal{R}}=(x\pm iy)/\sqrt{2}$
represent left and right hand polarization. For low frequency MHD
modes, we have from Ohm's Law $\mathbf{E}(\mathbf{k})=-(1/c)\mathbf{v}(\mathbf{k})\times\mathbf{B}_{0}$.
So we can express the electromagnetic fluctuations $T_{ij},R_{ij}$
in terms of correlation tensors $K_{ij}$. The particular dispersion relations for  $\omega$ are not important unless scattering happens at an angle close to $90^o$.For $\omega$ we use dispersion relations $\omega=k_\parallel V_A$ for Alfv\'en modes and $\omega=kV_{ph}$ for fast modes. Those dispersion relations were used to decompose and study the evolutions of MHD modes in CL02.

For gyroresonance, the
dominant interaction is the resonance at $n=\pm 1$. By integrating over
$\tau$, we obtain

\begin{eqnarray}
\left(\begin{array}{c}
D_{\mu\mu}\\
D_{\mu p}\\
D_{pp}\end{array}\right) & = & {\frac{\pi\Omega^{2}(1-\mu^{2})}{2}}\int_{\bf k_{min}}^{\bf k_{max}}dk^3\delta(k_{\parallel}v_{\parallel}-\omega+\pm \Omega)\left(\begin{array}{c}
\left(1+\frac{\mu V_{ph}}{v\zeta}\right)^{2}\\
\left(1+\frac{\mu V_{ph}}{v\zeta}\right)mV_{A}\\
m^{2}V_{A}^{2}\end{array}\right)\label{gyroapp}\\
 &  & \left\{ (J_{2}^{2}({\frac{k_{\perp}v_{\perp}}{\Omega}})+J_{0}^{2}({\frac{k_{\perp}v_{\perp}}{\Omega}}))\left[\begin{array}{c}
M_{{\mathcal{RR}}}({\mathbf{k}})+M_{{\mathcal{LL}}}({\mathbf{k}})\\
-C_{{\mathcal{RR}}}({\mathbf{k}})-C_{{\mathcal{LL}}}({\mathbf{k}})\\
K_{{\mathcal{RR}}}({\mathbf{k}})+K_{{\mathcal{LL}}}({\mathbf{k}})\end{array}\right]\right.\nonumber \\
 & - & 2J_{2}({\frac{k_{\perp}v_{\perp}}{\Omega}})J_{0}({\frac{k_{\perp}v_{\perp}}{\Omega}})\left.\left[e^{i2\phi}\left[\begin{array}{c}
M_{{\mathcal{RL}}}({\mathbf{k}})\\
-C_{{\mathcal{RL}}}({\mathbf{k}})\\
K_{{\mathcal{RL}}}({\mathbf{k}})\end{array}\right]+e^{-i2\phi}\left[\begin{array}{c}
M_{{\mathcal{LR}}}({\mathbf{k}})\\
-C_{{\mathcal{LR}}}({\mathbf{k}})\\
K_{{\mathcal{LR}}}({\mathbf{k}})\end{array}\right]\right]\right\} .\nonumber \end{eqnarray}

For TTD,

\begin{eqnarray}
\left(\begin{array}{c}
D_{\mu\mu}\\
D_{\mu p}\\
D_{pp}\end{array}\right) & = & {\frac{\pi\Omega^{2}(1-\mu^{2})}{2}}\int_{\bf k_{min}}^{\bf k_{max}}dk^3\delta(k_{\parallel}v_{\parallel}-\omega)\left(\begin{array}{c}
\left(1+\frac{\mu V_{ph}}{v\zeta}\right)^{2}\\
\left(1+\frac{\mu V_{ph}}{v\zeta}\right)mV_{A}\\
m^{2}V_{A}^{2}\end{array}\right)\label{TTD}\\
 &  & 2J_{1}^{2}({\frac{k_{\perp}v_{\perp}}{\Omega}})\left[\begin{array}{c}
M_{{\mathcal{RR}}}({\mathbf{k}})+M_{{\mathcal{LL}}}({\mathbf{k}})+e^{i2\phi}M_{{\mathcal{RL}}}({\mathbf{k}})+e^{-i2\phi}M_{{\mathcal{LR}}}({\mathbf{k}})\\
-C_{{\mathcal{RR}}}({\mathbf{k}})-C_{{\mathcal{LL}}}({\mathbf{k}})-e^{i2\phi}C_{{\mathcal{RL}}}({\mathbf{k}})-e^{-i2\phi}C_{{\mathcal{LR}}}({\mathbf{k}})\\
K_{{\mathcal{RR}}}({\mathbf{k}})+K_{{\mathcal{LL}}}({\mathbf{k}})+e^{i2\phi}K_{{\mathcal{RL}}}({\mathbf{k}})+e^{-i2\phi}K_{{\mathcal{LR}}}({\mathbf{k}})\end{array}\right]\nonumber \\
\nonumber \end{eqnarray}

Define the integrands 

\begin{eqnarray}
G(\mathbf{k})&= & (J_{2}^{2}({\frac{k_{\perp}v_{\perp}}{\Omega}})+J_{0}^{2}({\frac{k_{\perp}v_{\perp}}{\Omega}}))(K_{{\mathcal{RR}}}({\mathbf{k}})+K_{{\mathcal{LL}}}({\mathbf{k}}))-2J_{2}({\frac{k_{\perp}v_{\perp}}{\Omega}})J_{0}({\frac{k_{\perp}v_{\perp}}{\Omega}})(e^{i2\phi}K_{{\mathcal{RL}}}({\mathbf{k}})+e^{-i2\phi}K_{{\mathcal{LR}}}({\mathbf{k}}))\nonumber\\
T(\mathbf{k})&= & 2J_{1}^{2}({\frac{k_{\perp}v_{\perp}}{\Omega}})((K_{{\mathcal{RR}}}({\mathbf{k}})+K_{{\mathcal{LL}}}({\mathbf{k}})+2J_{2}({\frac{k_{\perp}v_{\perp}}{\Omega}})J_{0}({\frac{k_{\perp}v_{\perp}}{\Omega}})(e^{i2\phi}K_{{\mathcal{RL}}}({\mathbf{k}})+e^{-i2\phi}K_{{\mathcal{LR}}}({\mathbf{k}}))
\end{eqnarray}.
We shall show below how they can be simplified in various cases to
enable an analytical evaluation of the integral. The spherical components
of the correlation tensors are obtained in the following.

For Alfv\'en modes, their tensors are proportional to

\[
I_{ij}=\left(\begin{array}{cc}
\sin^{2}\phi & -\cos\phi\sin\phi\\
-\cos\phi\sin\phi & \cos^{2}\phi\end{array}\right).\]

Thus we get

\begin{eqnarray}
I_{{\mathcal{RR}}}=I_{{\mathcal{LL}}} =  {\frac{(I_{x}-iI_{y})}{\sqrt{2}}}{\frac{(I_{x}^{*}+iI_{y}^{*})}{\sqrt{2}}}\label{ARRLL}={\frac{1}{2}}(I_{xx}+I_{yy})={\frac{1}{2}},\nonumber\end{eqnarray}

and

\begin{eqnarray}
e^{i2\phi}I_{{\mathcal{RL}}}+e^{-i2\phi}I_{{\mathcal{LR}}} & = & {\frac{(I_{x}-iI_{y})^{2}}{2}}\times e^{i2\phi}+{\frac{(I_{x}+iI_{y})^{2}}{2}}\times e^{-i2\phi} = (I_{xx}-I_{yy})\cos2\phi+(I_{xy}+I_{yx})\sin2\phi\label{ARL}= -1.\nonumber
\end{eqnarray}

For Alfv\'en modes, $k_{\perp}v_{\perp}/\Omega\gg1$ because of the
anisotropy. So we can use the asymptotics of the Bessel function $J_{n}(x)=\sqrt{2/\pi x}\cos(x-n\pi/2-\pi/4)$
when $x\rightarrow\infty$. Thus from Eq.(\ref{anisotropic},\ref{ARRLL},\ref{ARL}),
we simplify the integrand in Eq.(\ref{gyro}), $G({\mathbf{k}})\propto k_{\perp}^{-16/3}\exp(-L^{1/3}|k_{\parallel}|/k_{\perp}^{2/3}),$
the integral of which will be $\propto(|k_{\parallel}|L^{1/3})^{-6.5}\Gamma[6.5,k_{max}^{-2/3}|k_{\parallel}|L^{1/3}]$.
Then from Eq.(\ref{gyro}) we can get the analytical result for the
scattering by Alfv\'en modes as given in Eq.(\ref{ana}).

For fast modes, their tensors have such a component

\begin{eqnarray}
H_{ij}=Ak^{-3.5}\left(\begin{array}{cc}
\cos^{2}\phi & \cos\phi\sin\phi\\
\cos\phi\sin\phi & \sin^{2}\phi\end{array}\right) &  & .\end{eqnarray}

Thus we have

\begin{eqnarray}
H_{{\mathcal{RR}}}=H_{{\mathcal{LL}}} & = & {\frac{1}{2}}(H_{xx}+H_{yy})=\frac{1}{2}\label{FRRLL}\end{eqnarray}

and

\begin{eqnarray}
e^{i2\phi}H_{{\mathcal{RL}}}+e^{-i2\phi}H_{{\mathcal{LR}}} & =(H_{xx}-H_{yy})\cos2\phi+(H_{xy}+H_{yx})\sin2\phi=1.\label{FRL}\end{eqnarray}

In general, it's more difficult to solve the integral in Eq.(\ref{genmu})
for the fast modes because they are isotropic. However, if taking
into account damping, Bessel function can be evaluated using the zeroth
order approximation. Thus from Eq.(\ref{gyro},\ref{FRRLL},\ref{FRL}),
we see for gyroresonance the integrand can be simplified as $G(\mathbf{k})\propto(J_{2}-J_{0})^{2}k^{-7/2}\propto k^{-7/2}$.
For TTD, $T(\mathbf{k})\propto J_{1}^{2}(k_{\perp}v_{\perp}/\Omega)k^{-7/2}\propto k^{-3/2}v_{\perp}^{2}\sin^{2}\theta\propto k^{-3/2}v_{\perp}^{2}(1-V_{ph}^{2}/v_{\parallel}^{2})$.
Then we can solve the integral in Eq.(\ref{TTD}) analytically and get the
scattering rate for fast modes.

\end{document}